# Improving Human-Computer Interaction by Developing Culture-sensitive Applications based on Common Sense Knowledge

Junia Coutinho Anacleto and Aparecido Fabiano Pinatti de Carvalho
*Advanced Interaction Lab / Computing Department / Federal University of São Carlos*
*Brazil*

## 1. Introduction

Computing is becoming ever more present in people's everyday life. Regarding this fact, researchers started to think of ways for improving Human-Computer Interaction (HCI) so that computer applications and devices provide users with a more natural interaction, considering the context which users are inserted into, as humans can do (Harper et al., 2008). Developing culture-sensitive interactive systems seems to be a possibility for reaching such goal. Bailey et al. (2001) already mentioned the importance of considering cultural issues in computer systems development. For them, culture is a system of shared meanings which forms a framework for solving problem and establishing behavior in everyday life that has to be considered in interactive system development. Since cultural knowledge is often implicit, designers often have trouble even realizing that their designs carry cultural dependencies implicitly. Moreover, it is not possible to design by hand for every combination of possible cultures, nor is it practical to exhaustively test for every possible user culture (Bailey et al., 2001). A support is necessary for making this possibility come true and information and communication technology can offer it, as it is explained further ahead.

This chapter presents the solutions of the Advanced Interaction Laboratory[1] (LIA) from the Federal University of São Carlos (UFSCar), Brazil, for developing culture-sensitive interactive systems. LIA's approach relies on using common sense knowledge for developing such kind of systems. That is because individuals communicate with each other by assigning meaning to their messages based on their prior beliefs, attitudes, and values, i.e. based on their common sense. Previous researches developed at the Lab have shown that common sense expresses cultural knowledge. So, providing this kind of knowledge for computers is a way of allowing the development of culture-sensitive computer applications (Anacleto et al., 2006a).

One idea for providing computers with common sense is to construct a machine that could learn as a child does, observing the real world. However, this approach was discarded after Minsky and Papert's experience of building an autonomous hand-eye robot, which should perform simple tasks like building copies of children's building-block structures. In this experience, they realized that numerous short programs would be necessary to give machines human abilities as cognition, perception and locomotion and that it would be very difficult to develop those programs (Minsky, 1988).

---

[1] LIA's homepage (in Portuguese): http://lia.dc.ufscar.br

Another idea is to build a huge common sense knowledge base, store it in computers and develop procedures that can work on it. This seems to be an easier approach; nevertheless there are at least three big challenges that must be won in order to achieve it. The first challenge is to build the necessary common sense knowledge base, since it is estimated that, in order to cover the human common sense, billions of pieces of knowledge such as knowledge about the world, myths, beliefs, and so on, are necessary (Liu & Singh, 2004). Furthermore it is known that common sense is cultural and time dependent, i.e. a statement that is common sense today may not be a common sense statement in the future (Anacleto et al., 2006b). For instance, consider the statement "The Sun revolves around the Earth". Nowadays this statement is considered wrong, however, hundreds of years ago people used to believe that it was right.

One possible idea to transpose this difficulty is to build the knowledge base collaboratively by volunteers through the Web, since every ordinary people has the common sense that computers lack (Liu & Singh, 2004, Anacleto et al. 2006a). In order to make the collection process as simple as possible to the volunteers, it is kind to think of collecting the common sense statements in natural language.

Then the second big challenge arises: to represent the knowledge collected in natural language in a way that computers can make inferences over it. In order to be used by computer inference mechanisms, it is still necessary that the knowledge be represented in specifics structures such as semantic network or ontology. So, it is necessary to process the statements in natural language in order to build a suitable knowledge representation. Natural language processing is a well-known AI challenge (Vallez & Pedraza-Jimenez, 2007).

The third challenge is to generate natural language from the adopted knowledge representation, so that computer systems can naturally and effectively communicate with users. This is another well-known challenge of current AI researches (Vallez & Pedraza-Jimenez, 2007).

This chapter discusses LIA's approaches for common sense knowledge acquisition, representation and use, as well as for natural language processing, developed in the context of the Brazilian Open Mind Common Sense (OMCS-Br) project (Anacleto et al., 2008b), in order to develop applications using such approach. It shows how common sense knowledge can be used for instantiating some issues of cultural-sensitive systems in the area of HCI. For this purpose, some applications developed at LIA are presented and the use of common sense knowledge in those applications is explained. Those applications are mainly developed for the domain of education, which is extremely culture-sensitive, and one of the main technological, social and economical challenges considering globalization and the necessary digital inclusion of every ordinary person.

The chapter is organized as follows: section 2 goes over some related projects that proposes to build large scale common sense knowledge bases such as Cyc (Lenat et al., 1990), the American Open Mind Common Sense (OMCS-US) and ThoughtTreasure (Mueller, 1997); section 3 presents details on OMCS-Br architecture for acquiring, representing, making available and using common sense knowledge in computer application; section 4 suggests ways of using common sense knowledge in the development of interactive systems and exemplifies the suggestions with computer applications developed at LIA; finally, section 5 presents some conclusions remarks and point to some future works.

## 2. Building Up Common Sense Knowledge Bases

Nowadays there are several projects, like Cyc, OMCS-US and ThoughtTreasure, which aim to build up large scale common sense knowledge bases. While those projects are mainly focused on Artificial Intelligence (AI) issues, OMCS-Br is more concerned about HCI. This section goes over the approach of such projects to build up large scale common sense knowledge bases, contextualizing LIA's researches.

In late 1950s, the goal of providing common sense knowledge for computers received great attention from AI researches. However, in the early 1960s that goal was abandoned by many of those researches, due especially to the difficulties they faced with for building the necessary common sense knowledge base (Minsky, 2000; Lenat, 1995).

Believing in the possibility of building up the knowledge base and making computer applications intelligent so that they could flexibly answer to users input through inferences mechanisms that would work on the common sense knowledge base, Douglas Lenat started in 1984 the Project Cyc, founding the Cycorp Inc. (Lenat et al., 1990). Nowadays, Cyc is one of the more active projects on acquiring and representing common sense knowledge computationally and on developing inferences mechanisms for common sense reasoning, as it can be verified in its scientific production[2].

At the beginning, the approach used by Cyc researchers was to construct the project knowledge base by hand. Panton et al. (2006) mention that the use of natural language understanding or machine learning techniques was considered by the project researchers, but it was concluded that, to have good results using those techniques, computers would have to have a minimum of common sense knowledge for understanding natural language and learning something from it. Therefore, it was defined a specific language for representing the knowledge, named CycL (Lenat et al., 1990), and, then, the corporation's *knowledge engineers*, trained in that language, started to codify and store pieces of knowledge in Cyc knowledge base. The knowledge was gotten from ordinary people by interviews about the several subjects that are part of people's common sense, in an effort of approximately 900 people per year (Matuszek et al., 2006).

Inspired in Cyc, Eric Mueller started the Project ThoughtTreasure in 1994, having as goal to develop a platform for making natural language processing and common sense reasoning viable to computer applications (Mueller, 1997). As in Cyc, the approach used for building ThoughtTreasure knowledge base was manual.

The existence of concepts, phrases and predicates both in English and French in ThoughtTreasure knowledge base is a differential of this project. The project also does not limit the knowledge representation to a logic language, as CycL does, but also allows knowledge representation in the format of finite automata and scripts (Mueller, 1997).

Besides the flexibility of ThoughtTreasure knowledge representation, it does not discard the need of knowing a specific structure to populate the base of the project. This issue, shared by Cyc and ThoughtTreasure projects, impedes that people who do not know the specific format that knowledge should be inserted into the knowledge base collaborate on building it. This demands a bigger time frame for making the knowledge base rise so that robust inferences can be done over it. Cyc and ThoughtTreasure are samples of this. After more than two decades, Cyc's knowledge engineers could map only 2.2 millions of assertions (statements and rules) (Matuszek et al., 2006). ThoughtTreasure, one decade younger, has only 50,000 assertions relating its 25,000 concepts (Mueller, 2007).

---

[2] See Cycorp Incorporation group's publications in http://www.cyc.com/cyc/technology/pubs

Thinking about fast building a large scale common sense knowledge base, researchers from the MediaLab of the Massachusetts Institute of Technology (MIT) launched OMCS-US, taking into account that every ordinary people have the common sense knowledge that machines lack, so all of them are eligible for helping build the common sense knowledge base machines need to be intelligent (Singh, 2002). In order to involve every ordinary people, OMCS-US collects common sense statements in natural language on a website made available by its researchers, and uses natural language processing techniques to build up its semantic network, the knowledge representation adopted in the project (Liu & Singh, 2004). Therefore, volunteers need only to know how to write in a specific language – nowadays there are three OMCS projects, each one in a language: OMCS-US, in English[3], OMCS-Br, in Portuguese[4] and OMCS-MX, in Spanish[5] – and have access to the Internet in order to collaborate in building the knowledge base. In less than five years OMCS-US got a semantic network with more than 1.6 millions assertions of common sense knowledge "encompassing the spatial, physical, social, temporal and psychological aspects of everyday life" (Liu & Singh, 2004). Section 3.1 gives details on the common sense collection process used by OMCS-Br, the same used by OMCS-US.

Being aware of the importance of including ordinary people in constructing its knowledge base, Cyc also started to develop systems for populating the knowledge base from natural language statements. The first try was made by developing KRAKEN (Panton et al., 2002). The system was developed in order to allow subject-matter experts to add information to pieces of knowledge already stored in Cyc knowledge base through natural language interface. In 2003, Witbrock et al. (2003) presented a system for interactive dialog, which used a structure similar to KRAKEN for allowing amateurs in CycL to enter information in the knowledge base.

Another approach for collecting common sense knowledge is proposed by von Ahn et al. (2006). Von Ahn's team agrees that collecting common sense knowledge in natural language is a good approach. However, they criticize the absence of fun in the collection process used by Cyc and OMCS-US, blaming this absence for until nowadays no project which has proposed to construct large scale common sense knowledge bases has beaten the amount of hundreds of millions facts that are supposed to be necessary to cover the whole human common sense. Thus, they propose to use games for fast collecting common sense facts, believing that through games it is possible to collect million of facts in few weeks (von Ahn et al., 2006). Considering this premise, the group developed Verbosity, a game in which common sense facts are collected as side effect of playing the game. In the same way, OMCS-US research group developed Common Consensus (Lieberman et al., 2007), aiming to motivate volunteers to collaborate in building the project knowledge base.

Last but not least, there is another approach for acquiring common sense knowledge under development at Cycorp, which aims to automate the collection process by using crawlers to get common sense knowledge from web pages. The developed method has six steps (Matuszek et al., 2005): *(i)* to use Cyc inference engine for selecting subjects of interest from Cyc knowledge base and composing query strings to be submitted to Google; *(ii)* to search for relevant documents on the Internet using Google's resources; *(iii)* to identify relevant statements to the subjects of interest on the retrieved documents; *(iv)* to automatically codify the statements in CycL; *(v)* to validate the new assertions using the knowledge already stored at Cyc common sense knowledge base and performing new searches on the Internet using Google; and *(vi)* to send the assertions to a knowledge engineer who decides whether they are going to be stored in the project knowledge base or not.

---

[3] OMCS-US Homepage: http://commons.media.mit.edu/en/
[4] OMCS-Br Homepage (in Portuguese): http://www.sensocomum.ufscar.br
[5] OMCS-Mx Homepage(in Spanish): http://openmind.fi-p.unam.mx

Although this last approach seems to be very innovative, it does not allow mapping the profile of the people from whom common sense knowledge was collected. OMCS-Br researchers defend that associating common sense knowledge to the profile of the person from whom it was collected is extremely important to make viable the use of such knowledge for HCI. This is because common sense knowledge varies from group to group of people and, therefore, the feedback that a common sense-aided application should give to users from different target groups should be different (Anacleto et al., 2006b). The variation in common sense among groups of people is what makes interesting use it for supporting the development of culture-sensitive computer application, as it is explained in section 4. That is why all OMCS-Br applications used for collecting common sense demands that users subscribe themselves and log into the system before starting interacting. In such way developers can analyze how people from different target groups talk about specific things and propose different solutions for each group.

Furthermore, the judgment performed by knowledge engineers makes possible question if the statements stored in Cyc knowledge base really represent common sense. Statements such as "Avocado is red" would not be stored in Cyc knowledge base, according to the process described by Matuszek et al. (2005), Shah et al. (2006) and Panton et al. (2006). However, if the majority of people with the same cultural background say that "Avocado is red", this should be considered as common sense for that community and should be in the knowledge base despite the knowledge engineers judge it as wrong. Because of that OMCS-Br does not judge the semantic of the statements collected in its site (Anacleto et al., 2008b).

## 3. The Project OMCS-Br

OMCS-Br is a project based on OMCS-US, which has been developed by LIA/UFSCar since August 2005. The project works on five work fronts: (1) common sense collection, (2) knowledge representation, (3) knowledge manipulation, (4) access and (5) use. Figure 1 illustrates the project architecture.

First of all, common sense knowledge is collected both through a website and through applications which have been developed in the context of OMCS-Br and stored in the OMCS knowledge base. After that, the knowledge is represented as semantic networks, called ConceptNets. In order to build the semantic networks, the natural language statements are exported through an **Export Module** and sent to the **Semantic Network Generator**. The generator is composed of three modules: the **Extraction, Normalization** and **Relaxation** modules. The result of these three modules is a group of binary predicates. For dealing with Portuguese natural language processing, Curupira (Martins et al., 2006), a natural language parser for Portuguese, is used. The predicates are filtered according to parameters of profile, such as gender, age and level of education, generating different ConceptNets. In order to make inferences over the ConceptNets, several inference mechanisms, which are grouped in an API (Application Programming Interface), were developed. The access to the API functions is made through a Management Server, which makes available access to instances of APIs associated with different ConceptNets. Through this access, applications can use the API inference mechanisms and can perform common sense reasoning. Details about each element of the architecture presented in Figure 1 and the OMCS-Br work fronts are presented in the following sub-sections.

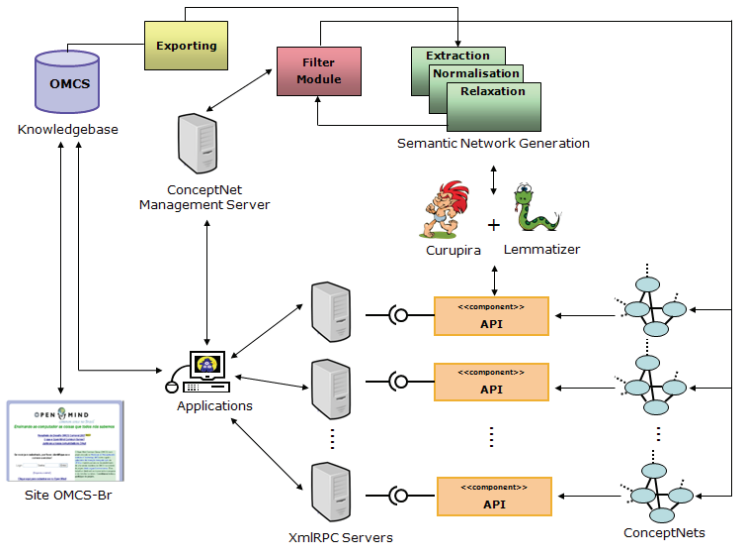

Fig. 1. OMCS-Br Project architecture

### 3.1. Common Sense Collection

OMCS-Br adopts template-based activities which guide users in such a way that they can contribute with different kinds of knowledge. The templates are semi-structured statements in natural language with some lacunas that should be filled out with the contributors' knowledge so that the final statement corresponds to a common sense fact. They were planned to cover those kinds of knowledge previously mentioned and to get pieces of information that will be used further to give applications the capacity of common sense reasoning. The template-based approach makes easier to manage the knowledge acquired, since the static parts are intentionally proposed to collect statements which can be mapped into first order predicates, which composes the project semantic networks. In this way, it is possible to generate extraction rules to identify the concepts present in a statement and to establish the appropriate relation-type between them. In OMCS projects, there are twenty relation-types, used to represent the different kinds of common sense knowledge, as it is presented in (Liu & Singh, 2004).

Those templates have a static and a dynamic part. The dynamic part is filled out by a feedback process that uses part of statements stored in the knowledge base of the project to compose the new template to be presented. Figure 2 exemplifies how the feedback process works. At the first moment the template "You usually find a __________ in a **chair**" of the activity Location is presented to a contributor – the templates bold part is the one filled out by the feedback system. In the example, the contributor fills out the sentence with the word "screw". Then, the sentence "You usually find a **screw** in a chair" is stored in the OMCS knowledge base. At the second moment, the template "A screw is used for __________" of the activity Uses is shown to another contributor. Note that the word "screw" entered at the first moment is used to compose the template presented at the second moment.

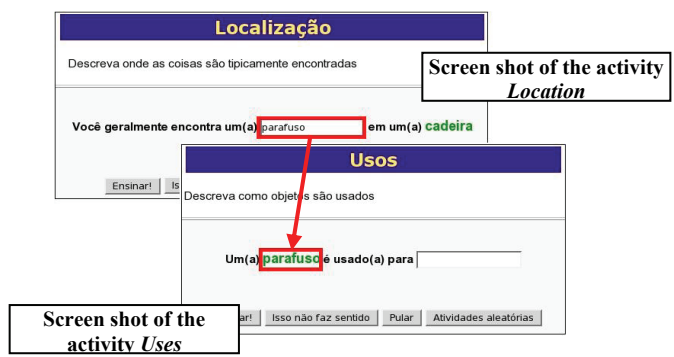

Fig. 2. Example of the OMCS-Br feedback process

The feedback process used in OMCS-Br website was planned in order to allow varied templates to be generated so that users are able to contribute on several subjects and do not get bored with always filling out the same sentence.
Still related to the feedback process, considering that the statements stored in the knowledge base will be used to compose templates that will be shown to other contributors, it is important to provide a way through what it could be selected the statements that should be used by the feedback process. Thinking in this need, it was developed in OMCS-Br an on-line review system, which can be accessed just by the ones who have administrator privileges, where the statements are selected to be or not to be used by the feedback process. In order to perform the review, it was defined some rules to assure that common sense knowledge would not be discarded (Anacleto et al., 2008b). It is worth pointing out that during the review process the reviewer is not allowed to judge the semantic of a sentence. That is because it does not matter if a sentence seems strange in meaning or if it has already been scientifically proved as wrong. Common sense knowledge does not match scientific knowledge necessarily. Since a sentence is accepted as true by the most people who share the same cultural background, it is considered as a common sense sentence. Because of that , reviewers are not allowed to judge if a sentence is common sense statement or not. Only statements with misspelling errors are discarded, in order not to cause noisy in the semantic networks generated in the next step (Anacleto et al., 2008b).
Besides the templates about general themes such as those about "things" which people deal with in their daily life, "locations" where things are usually found and the common "uses" of things, there are also, in the Brazilian project website, templates about three specific domains: health, colors and sexual education. They are domains of interest to the researches that are under development in the research group which keeps the project (Anacleto et al., 2008b). This approach is only used in Brazil and it was adopted taking into account the necessity of making the collection of common sense knowledge related to those domains. The specific-domain templates were defined with the help of professionals of each domain. They were composed with some specifics words which instantiate the templates of general themes in the domain, in order to guide users to contribute with statements related to it. Table 1 shows the accomplishments that OMCS-Br has gotten with that approach.

| domain | number of contributions | period of collection |
| --- | --- | --- |
| Healthcare | 6505 | about 29 months |
| Colors | 8230 | about 26 months |
| Sexual Education | 3357 | about 21 months |

Table 1. Contributions on specific domains in OMCS-Br

The numbers of contributions in each domain can seem to be irrelevant, however, considering the only 2 statements about AIDS found in the knowledge base before creating the theme Sexual Education, it can be noticed the importance of domain-contextualized templates in order to make faster the collection of statements related to desired domains.

Another accomplishment of the OMCS-Br is related to the variety of contributor profiles. Nowadays there are 1499 contributors registered in the project site of which 19.33% are women and 80.67% are men. The most part of contributors (72.80%) is from Brazil South-east area, followed by the South area (15.25%). Those numbers point to the tendency proved by geographic sciences, which present the South-east and South area as being the most developed areas of Brazil. Considering that, it is perfectly understandable that, being well developed areas, their inhabitants have easier access to the Internet. Table 2 and Table 3 present other characteristics of OMCS-Br contributors.

| age group | percentage |
| --- | --- |
| Younger than 12 years | 0.75 % |
| 13 – 17 | 20.51 % |
| 18 – 29 | 67.36 % |
| 30 – 45 | 9.88 % |
| 46 – 65 | 1.22 % |
| Older than 65 years | 0.28 % |

Table 2. Percentage of Contributors by Age Group

| school degree | percentage |
| --- | --- |
| Elementary school | 2.21 % |
| High school | 18.17 % |
| College | 65.86 % |
| Post-Graduation | 4.52 % |
| Master Degree | 7.04 % |
| Doctorate Degree | 2.21 % |

Table 3. Percentage of Contributors by School Degree

Another conquest of OMCS-Br is the amount of contributions. Within two years and a half of project, it has been gotten more than 174.000 statements written in natural language. This was possible thanks to the web technology and the marketing approach adopted by LIA. As the project was released in Brazil in 2005, it was realized that the knowledge base would rise up significantly just when there were an event that put the project in evidence. Figure 3 demonstrates this tendency.

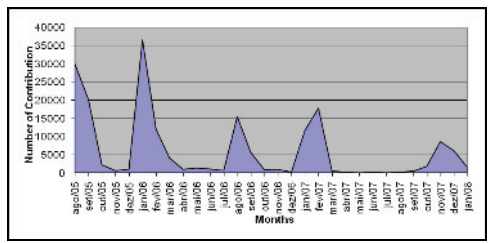

Fig. 3. OMCS-Br knowledge base tendency of growing up

It can be noticed in Figure 3 that the periods where the knowledge base grew up significantly were from August to October 2005, from January to March 2006, from August to October 2006, from January to February 2007 and from November to December 2007. This is an interesting fact, because those jumps in the knowledge base just followed some marketing appeals performed by LIA. In the first one, LIA got published some articles in newspapers of national coverage telling people about the project and asking for people contribution. After getting those articles printed, the OMCS-Br knowledge base reached the number of 50.000 statements. Three months later, the knowledge base established and passed to grow up very slowly.
Thinking of having another jump in the knowledge base size, it was released in the later January 2006 a challenge associated to the Brazilian carnival. In that challenge, it was offered little gifts as prizes to the three first collaborators that contributed with the most number of statements in the site activities. The winners received T-Shirts of the OMCS-Br Project and pens of MIT. The challenge was announced among the project contributors, who received an e-mail telling about it. The announcement was also posted in the Ueba website (www.ueba.com.br), a site of curiosities which target public is people interested in novelties. As it can be noticed, the knowledge base size had a jump as soon as the challenge was launched. The same approach was used in August 2006, January 2007 and December 2007.
Although the approach has gotten a good response from the contributors in the first three challenges, it can be noticed in Figure 3 that this approach is becoming inefficient. Thinking about keeping the knowledge base growing up, it is under development some games, following project contributors' suggestions, in order to make the collection process funnier and more pleasant.
One example is the game framework "What is it?" (Anacleto et al., 2008a), presented in section 4.3. The framework is divided in two modules: (1) the player's module, a quiz game where players must guess a secret word considering a set of common sense-based clues and (2) the game editor's module, a seven-step wizard which guides people to create game instances by using common sense knowledge. In the player's module, while the player tries to find out the secret word, the system collects common sense knowledge storing the relationship between the word suggested and the clues already seen by the player. In the game editor's module common sense knowledge is collected during the whole set up process, where the editor defines: (a) the community profile whose common sense statements should be considered; (b) the game main theme; (c) the topics, i.e. specifics subjects related to the main theme; and (d) the cards of the game. All possible combinations among the data supplied by the system and the editor's choices are stored in the knowledge base. For example, the secret word select by the editor and the synonyms (s)he are stored in the knowledge base as statements like "*secret word is also known as synonym*". Moreover, statements relating the synonyms among them, i.e. statements like "*synonym-1 is also known as synonym-2*",…, "*synonym-(n-1) is also known as synonym-n*", are stored.

### 3.2. Knowledge Representation

As OMCS-Br adopts natural language approach to populate its knowledge base, it was decided to pre-process the contributions stored in the knowledge base so that they could be easier manipulated by inference procedures. From this pre-processing, semantic networks are generated with the knowledge represented as binary relations. These semantic networks are called ConceptNets in OMCS projects.

Currently twenty relation-types are used to build the OMCS-Br ConceptNets. Those relation-types are organized in eight classes, presented by Liu and Sing (2004). They were defined to cover all kinds of knowledge that compose the human common sense – spatial, physical, social, temporal, and psychological (Liu & Singh, 2004). Some of the relation-types, the K-Lines, were adopted from Minsk's theory about the mind. In that theory, K-Line is defined as a primary mechanism for context in memory (Minsky, 1988).

ConceptNet relations have the format:

*(Relation-type "parameter-1" "parameter-2" "f=;i=" "id_numbers")*

Figure 4 depicts the graphical representation of a mini ConceptNet. That semantic network was generated from a very small excerpt of the OMCS-Br knowledge base. The nodes, originally in Portuguese, was freely translated for this chapter.

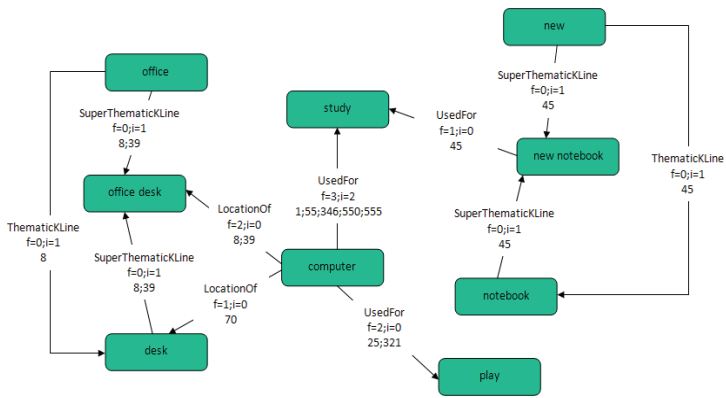

Fig. 4.  Sample of ConceptNet

One example of relation that can be mapped from Figure 4 is:

*(UsedFor "computer" "study" "f=3;i=2" "1;55;346;550;555")*

The parameters *f* (**f**requency of uttered statements) and *i* (frequency of **i**nferred statements) are part of the semantic relations. The *f* represents how many times a relation was generated directly from the statements stored in the knowledge base, through extraction rules. The *i* represents how many times a relation was generated indirectly from the statements. In this case they are submitted to the natural language parser and afterwards inference heuristics are applied on them. The generation of the *f* and *i* is explained further ahead. The *id* numbers correspond to the *id* of the natural language statements in the knowledge base which originated the relation. This allows computer programs to go back to the knowledge base and retrieve natural language statements related to a certain context of interaction.

In OMCS-Br, differently from OMCS-US, it is possible to generate several ConceptNets considering only the statements collected from a specific profile of volunteers. This feature was incorporated in the Brazilian project since it is known that common sense varies from group to group of people, inserted in different cultural context (Anacleto et al., 2006a; Carvalho et al., 2008). The profile parameters which can be used to filter the knowledge from the knowledge base and to generate different ConceptNets are: age, gender, level of education and geographical location.

Furthermore, in OMCS-Br, each non k-line relation-type has its negative version. For example, in OMCS-Br ConceptNets, it can be found both "IsA" and "NotIsA" relations, "PropertyOf" and "NotPropertyOf", and so on. The decision for the affirmative or negative relation version is made in the second phase of the ConceptNet generation process. In total there are 5 phases in the ConceptNet generation process (Carvalho, 2007): (1) **Exporting,** (2) **Extraction,** (3) **Normalization,** (4) **Relaxation** and (5) **Filtering**. Each step is presented in details in the following.

### *Exporting*

The first step to generate an OMCS-Br ConceptNet is to export the natural language statements collected on the site and stored in the project knowledge base. This export is made by a PHP script that generates a file whose lines are seven-slot structures with the statements and information about them. The slots are separated one another through "$$". Figure 5 presents a very small excerpt of the OMCS-Br knowledge base after the export phase. Although the data in Figure 5 and in the other Figures related to the ConceptNet generation process are in Portuguese, they are explained in English in the text.

| |
|---|
| Um(a) computador é usado(a) para estudar$$M$$18_29$$mestrado$$Clementina$$SP$$1 |
| Um(a) computador é usado(a) para jogar$$M$$13_17$$2_incompleto$$São Carlos$$SP$$25 |
| Você geralmente encontra um(a) computador em uma mesa de escritório$$F$$18_29$$mestrado$$Campinas$$SP$$8 |
| Pessoas usam cadernos novos quando elas começam a estudar$$M$$13_17$$2_incompleto$$São Carlos$$SP$$45 |

Fig. 5. Sample result of the exporting phase

As it can be seen in Figure 5, the slot 1 stores the natural language statement, the slots 2 to 6 present information about the profile of the volunteer who entered the statement in the knowledge base (gender, age group, level of education, origin city and origin state, respectively) and the last slot stores the *id* number of the sentence in the knowledge base. For example, in the first structure presented in Figure 5: *(i)* the natural language statement is "Um(a) computador é usado(a) para estudar" (in English, "A(n) computer is used for studying"); *(ii)* the profile parameters are "male" (M), "person between 18 and 29 years old" (18_29), "Master of Science" (mestrado), "Clementina City" (Clementina), "São Paulo State" (São Paulo); and the knowledge base statement *id* number is "1". The id is very important for common sense-based applications to trace back the natural language statements related to certain interaction scenarios. This is better explained in section 3.3.

***Extraction***

In the extraction phase, extraction rules, i.e. regular expression patterns (Liu & Singh, 2004), are used for identifying specifics structures in the natural language statements. These regular expressions are designed according to the templates of the OMCS-Br site. Based on the templates, it is possible to identify the relation-type and the parameters of a relation. For example, considering the first statement of Figure 5, "Um(a) computador é usado(a) para estudar", through a extraction rule it is possible to identified the following structures "computador" ("computer"), "é usado(a) para" ("is used for"), "estudar" ("studying"). By these structures it is generated the relation (UsedFor "computador" "estudar"). Figure 6 shows the relations generated from the sample structures presented in Figure 5. It can be noticed in Figure 6 that the profile parameter are still in the relation structure. Those parameters are kept until the Filtering phase, where they are used to generate different ConceptNets, according to the application needs.

---

(UsedFor "computador" "estudar" "M" "18_29" "mestrado" "Clementina" "SP" "1")
(UsedFor "computador" "jogar" "M" "13_17" "2_incompleto" "São Carlos" "SP" "25")
(LocationOf "computador" "mesa de escritório" "F" "18_29" "mestrado" "Campinas" "SP" "8")
(MotivationOf "usam cadernos novos" "começam a estudar" "M" "13_17" "2_incompleto" "São Carlos" "SP" "265")

---

Fig. 6. Result of the Extraction phase for the structures in Figure 5

In this phase, it is decided which version of the relation-type should be used, the affirmative or the negative. The extraction rules decide for one of them taking into account the following heuristic: "*If there is a negative adverb before the structure which defines which relation-type should be used, then use the negative version. Use the affirmative version instead, if no negative adverb precedes the structure which defines the relation-type*".

For example, consider the following natural language statement: "Você quase nunca encontra um(a) mesa de escritório em um(a) rua" ("You hardly ever find a(n) office desk in a(n) street"). After being processed by an extraction rule, the structures "quase nunca encontra" ("hardly ever find"), "mesa de escritório" ("office desk") and "rua" ("street") are identified. In this case the verb "encontra" ("to find") expresses the semantics that leads to the relation-type "LocationOf". However, as the adverbial phrase "quase nunca" ("hardly ever") precedes the verb, the generation system decides to use the negative version of the relation-type. So, it is generated the relation (NotLocationOf "mesa de escritório" "rua").

The incorporation of negative version to the ConceptNet relation-types was an initiative of the OMCS-Br. Until 2007, the OMCS-US did not take into account this feature, and after some discussions led by the Brazilian research group, the American research group implemented the polarity parameter in their ConceptNet (Havasi et al., 2007). Nonetheless the Brazilian group decided to define new relation-types instead of implement another parameter in the ConceptNet relations, since this approach would cause a smaller impact in the inferences procedures which had been already developed.

*Normalization*

Since the statements collected in the site can express the same knowledge about the human aspects in several ways, such as, using synonyms, different phrase structures, and different inflectional morphology, those statements should be manipulated to increase the semantic network quality. In order not to have inflected concepts, which means the same word varying in number, tense, etc., a normalization process is performed.

The normalization process of OMCS-US makes use of Montylingua (Liu & Singh, 2004) – a natural language parser for English, that tags and strips inflectional morphology of the knowledge base statements. However, Montylingua cannot be used to parse Brazilian Portuguese statements, due to syntactic and lexical differences between those languages.

The alternative found by OMCS-Br was to adopt Curupira (Martins et al., 2003), a parser developed for Brazilian Portuguese. However Curupira does not strip the sentence inflectional morphology. Because of that, a module to normalize the statements submitted to OMCS-Br generation process was developed, using the inflectional dictionary developed by the UNITEX-PB project (UNITEX-PB, 2008), which has all inflectional forms of Brazilian Portuguese morphological classes. The "normalizer" works in 3 steps, as it is described in the following (Anacleto et al., 2008):

1. First of all, each sentence token is tagged using Curupira. Here it was necessary to develop another module which makes the bridge between the "normalizer" and the parser library, available through a *dll* (dynamic linked library) format.
2. Afterward, articles are taken off – proper names are kept in original form. Special language structures that are proper of Brazilian Portuguese are treats. For instance, the structure called *ênclise*, a case of pronominal position where the pronoun goes after the verb, is stripped from the statements and the verb is put in the infinitive form. For example, the verb "observá-la" ("observe it") is normalized to "observar" ("to observe").
3. Overall, each tagged token is normalized into its normal form found in the adopted inflectional dictionary. In this way, the statements that were separated by morphological variations, like "compraria cadernos novos" ("would buy new notebooks") and "comprou um caderno novo" ("the bought a new notebook"), are reconciled during the normalization process generating the normalized expression "comprar caderno novo" ("to buy new notebook").

For the purpose of measuring the effects of the normalization process on the semantic network, the average nodal edge-density, as well as the number of distinct nodes and of distinct relations in OMCS-Br ConceptNet, was calculated. This processing was performed using and not using the normalization process. The results of this measurement are presented in Table 4.

|  | non-normalized | normalized | normalized/ non-normalized |
|---|---|---|---|
| **nodes** | 36,219 | 31,423 | - 13.24 % |
| **relations** | 61,455 | 57,801 | - 5.95 % |
| **average nodal edge-density** | 4.4643 | 3.3929 | + 31.57 % |

Table 4. Effects of the normalization process on the Brazilian ConceptNet structure

It can be seen in the results that the number of nodes and relations were decreased after the normalization process. This confirms the tendency that the normalization process makes reconciliations between morphological variations, and thus unifies them.

Another result that can be inferred examining the connectivity of semantic network is that the nodal edge-density has increased more than 30%. This is the most relevant data of the measurement performed, since it demonstrates that the normalization process improves the connectivity of the nodes.

It is worth mentioning that the OMCS-Br research group has been facing with several difficulties regarding the use of Curupira. For instance, when an *ênclise* occurs, Curupira assigns more than a tag to the same token. The same happens with some composite words such as "*fada-madrinha*" ("fairy godmother") and "*cavalo-marinho*" (hippocampus). Moreover, some verbal phrases are tagged incorrectly. For instance, "*fazer compras*" (go shopping) is tagged as "fazer/VERB" and "compras/VERB" when it should be tagged as "fazer/VERB" and "compras/SUBST". However, some heuristics to transpose these difficulties has been developed, as it can be verified in (Carvalho, 2007).

### *Relaxation*

Other strategy developed to improve the ConceptNet connectivity is to extract new relations from the relations uttered directly from the natural language statements. This is made applying a set of heuristic inferences over the relations generated in the Extraction phase. The heuristics applied in this phase are based on grammatical and semantic patterns, as it is explained in the following (Carvalho, 2007).

The Relaxation module receives as input a file with the relations generated in the first phase of the ConceptNet generation process normalized and tagged, as it can be notice in Figure 7.

---

(UsedFor "computador/SUBST" "estudar/VERB" "M" "18_29" "mestrado" "Clementina" "SP" "1")

(UsedFor "computador/SUBST" "jogar/VERB" "M" "13_17" "2_incompleto" "São Carlos" "SP" "25")

(LocationOf "computador/SUBST" "mesa/SUBST de/PREP escritório/SUBST" "F" "18_29" "mestrado" "Campinas" "SP" "8")

(MotivationOf "usar/VERB caderno/VERB novo/ADJ" "começar/VERB a/PREP estudar/VERB" "M" "13_17" "2_incompleto" "São Carlos" "SP" "265" )

---

Fig. 7. Sample result of the Normalization phase

As an example, consider the last relation in the group. In Figure 6, that relation was like:

(MotivationOf "usam cadernos novos" "começam a estudar" "M" "13_17" "2_incompleto" "São Carlos" "SP" "265")

After the normalization process, the relation got the following representation:

(MotivationOf "usar/VERB caderno/VERB novo/ADJ" "começar/VERB a/PREP estudar/VERB" "M" "13_17" "2_incompleto" "São Carlos" "SP" "265" )

As it can be noticed, the verbs "usam" ("to use") e "começam" ("to start") were put in the infinitive form and tagged as verb (VERB); the noun "cadernos" ("notebooks") was put in singular and tagged as substantive (SUBST); the adjective "novos" was put in singular, since in Portuguese adjectives have also plural form, and was tagged as adjective (ADJ); and the preposition "a" received the preposition tag (PREP). The tags are very important in this phase because they are used by the inference mechanisms as it is explained in the following. The first step in the relaxation phase is to assign the parameters $f$ and $i$ to the relations. All the relations receives "f=1; i=0" at the beginning, because they are generated just once by an extraction rule and up to this point no relations were generated by inference mechanisms. The second step is to group all equal relations, incrementing the f parameter and appending the id number. For example, consider the following two relations:

(UsedFor "computador/SUBST" "jogar/VERB" "M" "13_17" "2_incompleto" "São Carlos" "SP" "25" "f=1;i=0")

(UsedFor "computador/SUBST" "jogar/VERB" "M" "13_17" "2_incompleto" "São Carlos" "SP" "387" "f=1;i=0")

They were generated from two different natural language statements (the statements number 25 and 387). However, they are the same relation and it was collected from people with the same profile (male, between 13 and 17, high school, São Carlos, SP). Note that although the profile parameters are the same, this does not mean that the statements 25 and 387 were collected from the same person, but from people with the same profile. After this second step, the two relations are reconciled in the following relation:

(UsedFor "computador/SUBST" "jogar/VERB" "M" "13_17" "2_incompleto" "São Carlos" "SP" "25;387" "f=2;i=0")

Note that *f* received a new value (f=2) and the id numbers were groped ("25;387"). If ten people with the same profile had entered a statement which generated that relation, *f* would be "f=10" and there would be 10 id numbers in the id number slot.
The next step of the relaxation phase is to generate new "PropertyOf" relations. They are generated from "IsA" relations. All IsA relation whose first parameter is a noun or a noun phrase and the second parameter is an adjective, generates a new "PropertyOf". For example the relation:

(IsA "computador/SUBST pessoal/ADJ" "caro/ADJ" "M" "18_29" "2_completo" "São Carlos" "SP" "284" "f=1;i=0")

generates the relation:

(PropertyOf "computador/SUBST pessoal/ADJ" "caro/ADJ" "M" "18_29" "2_completo" "São Carlos" "SP" "284" "f=0;i=1")

Note that the profile parameters and the id number are kept the same as the relation used by the inference process. It is worth pointing out that for each new relation generated, it is verified whether an equal relation is already in the ConceptNet. In case affirmative the *i* parameter of the existing relation is incremented and the id number of the generated relation is appended to its id numbers. For instance, consider that when the relation previously presented is generated, there is already the following relation in ConceptNet:

(PropertyOf "computador/SUBST pessoal/ADJ" "caro/ADJ" "M" "18_29" "2_completo" "São Carlos" "SP" "45;78;171" "f=3;i=0")

Instead of registering the generated relation in ConceptNet as a new relation, the existing relation is updated to:

(PropertyOf "computador/SUBST pessoal/ADJ" "caro/ADJ" "M" "18_29" "2_completo" "São Carlos" "SP" "45;78;171;284" "f=3;i=1")

Notice that the parameter *i* is now 1 and the *id* number 284 is part of the relation id numbers. New relations "CapableOf","CapableOfReceiveingAction","ThematickKLine" and "SuperThematicKLine" are created by similar processes. For detail about the other inference mechanisms that generate such relations, see (Anacleto et al., 2008b).

*Filtering*
After the Relaxation phase, different ConceptNets can be generated, according to the possible combination of the profile parameter values. This generation is made on demand, as common sense based applications which use the OMCS-Br architecture need a certain ConceptNet. This is only possible in OMCS-Br, since the OMCS-US, Cyc and ThoughtTreasure do not register their volunteers' profile.
The Filtering module receives an array of arrays as input with the profile parameter values which should be considered to generate the desired ConceptNet. The first sub-array in the global array has the parameters related to the volunteers' gender; the second, to their age group; the third, to their level of education; the forth, to the city they come from; and the fifth, to the state they live. If an array in the global array is empty, it means that a specific profile parameter does not matter for the desired ConceptNet, and then it is considered all possible value for that parameter. For example, if the array *[[], [13_17, 18_29], [2_completo], [], [SP, MG]]* is provided to the Filtering module, a ConceptNet will be generated, whose relations were build from statements gotten from people of both gender, since the first sub-array is empty; who are between 13 and 17 years old and between 18 and 29 years old, whose highest level of education is high school; who come from any city located in the Brazilian *São Paulo* and *Minas Gerais states*.
The first step in the Filtering phase is to recalculate the *f* and *i* parameter values, grouping the equal relations whose profiles fit to the profile values previously provided. After that, another heuristic is applied on the relations in order to generate new "PropertyOfs". This heuristic is applied only in this step because it considers two groups of relations in the inference process and these two groups should have only relations which fit to the considered profile. Therefore, for guaranteeing this constraint, it was decided to apply this heuristic only in this stage. For details about this heuristic, see (Carvalho, 2007).
After the Filtering phase, the ConceptNet is stored in a ConceptNet Server so that it can be accessed by the ConceptNet Server Management, as it is explained in details in section 3.5.

### 3.3. Knowledge Manipulation

Once the semantic network is built, it is necessary to develop procedures to manipulate it, in order to make computer applications capable of common sense reasoning and, then, to use these resources for developing culture-sensitive computer applications. The procedures developed in the context of OMCS-Br are integrated in its API for being used by computer applications. Currently there are five basic functions that simulate some sorts of human reasoning ability. They are related to context, projections, analogy making, topic gisting and affective sensing (Liu & Singh, 2004). The OMCS-Br applications use mainly three of them.

The first one is the *get_context()* function, which determines the context around a certain concept or around the intersection of several concepts. For example, when someone searches for the concept "have lunch" using *get_context()*, it returns concepts like "be hungry", "eat food", "meet friend", "buy food", "take a break", and "find a restaurant". These related concepts are retrieved by performing spreading activation from the source node in the semantic network that finds related concepts by and considering the number and the intensity of connected pairs of concepts. This function is very useful for semantic query expansion and topic generation. For instance, in the "What is it?" framework (Anacleto et al., 2008a), it is important to expand an initial concept provided by the game editor in order to bring more relations about the main concept.

The second function is *display_node()*, that can be used for bringing relations about a particular concept. This function retrieves: the relation-type, the two concepts associated through it, and the id numbers of the statements in knowledge base which originated the relation. Therefore, the applications can use these results to create complete natural language statements. Basically, there are two ways of doing so: (1) the relations can be mapped in statements such as the templates used in the OMCS-Br site; for example, the relation (UsedFor "computer" "study") enable to create a statement like "A computer is used for study"; (2) the id numbers can be used to find the original statements in the knowledge base. Note that in the first case there are still difficulties to generate natural language statements grammatically correct. This is one of the AI challenges concerning natural language generation (Vallez & Pedraza-Jimenez, 2007).

Other function is *get_analogy()*, which has been developed based on the Gentner's Theory of Mapping of Structures (TME) (Gentner, 1983). From the Structure-Mapping Engine (SME) (Falkenhainer et al., 1990), an algorithm capable of generating analogies and similarities from OMCS-Br ConceptNets was developed. This algorithm uses a ConceptNet as basis domain and an ExpertNet as target domain, returning literal similarities. Details about the algorithm can be found in (Anacleto et al., 2007b). In the same way it can be found details about the other common sense inference procedures used in OMCS-Br in (Liu & Sing, 2004).

### 3.4. Access

The ConceptNet API is available to any computer application through XML-RPC (Remote Procedure Call) technology (XML-RPC, 2008), which allows a simple connection between a client application and a server application over the Internet. This connection is established through HTTP and all data are encoding in XML format. First of all, the application informs the ConceptNet Management Server shown in Figure 1 about the profile parameters values related to the desired ConceptNet. The server checks whether an API for that ConceptNet has been already instantiated in a specific port handled by it. If the desired ConceptNet API has been not been instantiated yet, the server asks for the Filtering module to verify whether the desired ConceptNet has been already generated in another moment so that it can instantiate an API for it. In case affirmative, the server allocates a port for the ConceptNet and makes it available also through the XML-RPC protocol. If the ConceptNet has not been generated yet, the Filtering module generates it, and then the server instantiates and makes available an API for the ConceptNet so that the application can use the API inference procedures. Since the application has been informed about the port where the ConceptNet API is available, it can connect to the port and use the API procedures.

## 4. Designing and Developing Culture-Sensitive Applications

As the complexity of computer applications grows, one way to make them more helpful and capable of avoiding unwise mistakes and unnecessary misunderstandings is to make use of common sense knowledge in their development (Lieberman et al., 2004). LIA has been experiencing that cultural differences registered in common sense knowledge bases can be helpful in: (a) developing systems to support decision-making by presenting common sense knowledge of a specific group to users; (b) developing systems capable of common sense reasoning, providing different feedback for people from different target group; and (c) helping designers who want to consider these differences in the interactive systems development by customizing interfaces and content according to the user's profile.

In the following, those three items are approached in details considering the domain of education and examples of cultural sensitive common sense-aided computer applications for each of them are presented.

### 4.1. Developing systems that show cultural issues for decision-making

There are situations in which it is interesting to know the common sense of a specific group in order to make suitable decisions. This is especially interesting in the domains of Human-Human Interaction (HHI), when two or more people from different cultural background are interacting with each other, and of Education, where it is important for educators to know how their learners talk about themes which is going to be taught, so that they can decide on how to approach those themes during the learning process. This was the first kind of common sense-aided computer application development approached by LIA's researchers.

In the following, two applications developed at LIA to illustrate how common sense knowledge can be used for this purpose are presented.

#### *WIHT: a common sense-aided mail client for avoiding culture clashes*

As the world economy becomes more and more globalized, it is common to see situations where people from two or more cultural backgrounds have to communicate with each other. Developing communication systems which show cultural differences to people who want to communicate with each other, making commentaries on the differences in the grounding that can lead to possible misunderstandings so that someone can correct him/herself before getting into an embarrassing situation, can be considered an advance in HCI that reflects on HHI. This issue has been approached by the development of an e-mail client which shows cultural difference between people from three different cultures combined two by two: Brazilian, Mexican and American (Anacleto et al., 2006a).

The application has an agent that keeps watching what the user is typing, while makes commentaries on the differences in the grounding that can lead to possible misunderstandings. The system also uses these differences to calculate analogies for concepts that evoke the same social meaning in those cultures. This prototype is focused on the social interaction among people in the context of eating habits, but it could scale to other domains. The system interface has three sections, as can be seen in Figure 8. The first one – at the upper left – is the information for the e-mail addresses and the subject; the second one – at the upper right – is where the agent posts its commentaries about the cultural differences and the third part – the lower part – is the body of the message. The second section has four subsections: the upper one shows the analogies that the agent found and the other three show the data that are not suitable for analogy. For example, in the screen shot in figure 8, the third label for the Mexican culture – Mexicans thinks that dinner is coffee and cookies – and the second for American culture – Americans think that dinner is baked chicken – cannot make a meaningful analogy even if they differ only in one term.

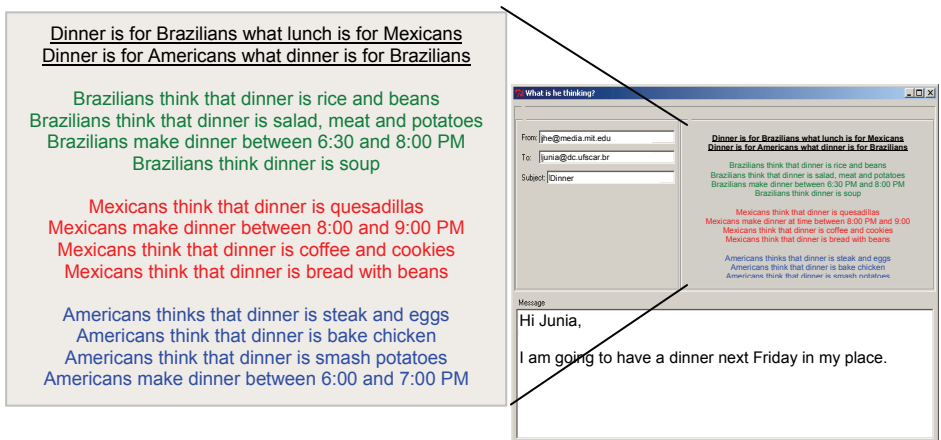

Fig. 8. WIHT screen shot (Anacleto et al., 2006a, p. 7)

In order to make the cultural analogies, the system uses three culturally specific semantic network that have knowledge about the Brazilian, Mexican and North-American culture – the ConceptNetBR, ConceptNetMX and ConceptNetUS respectively. The ConceptNetBR was built from data mined from the OMCS-Br knowledge base, originally in Portuguese. Specifically in this project, a small group of statements related to eating habits were selected and freely translated to English to be parsed by the system (Anacleto et al., 2006a).

### *PACO-T: a common sense-aided framework for planning learning activities*
Another application developed at LIA for supporting decision making is PACO-T (Carvalho et al., 2008b), a computational tool for PACO (**P**lanning learning **A**ctivities based on **CO**mputers), a seven-step framework which aims to support educators in planning pedagogically suitable learning activities (Neris et al., 2007). PACO seven steps are: (1) define the learning activity theme, target public and general goal; (2) organize the learning activity topics; (3) choose a pedagogical/methodological reference; (4) plan the learning tasks; (5) choose computer tools to support the tasks execution; (6) edit the learning objects which are going to be used in the learning activity; and (7) test pedagogical and technological issues.

Researches developed at LIA shows that common sense knowledge can be used for educators to (Carvalho et al., 2008a) *(i)* identify topics of general interest to the target group; *(ii)* fit the learning activity content to the target group's previous knowledge; *(iii)* identify suitable vocabulary to be used in the learning activity; and *(iv)* identify knowledge from the target group's domain to which new knowledge can be anchored so that meaningful learning can be achieved (this knowledge can be used, for instance, in composing metaphors and analogies to be presented during the learning activity). Those are some pedagogical issues necessary to allow effective learning to take place, according to Learning Theories from renowned authors from the field of Pedagogy, such as Freire (2003), Freinet (1993), Ausubel (1968) and Gagné (1977).

Therefore, a case study was conducted specially to check the possibility of using common sense knowledge during the planning of learning activities by using PACO, to support teachers in answering questions brought up along the framework steps. The case study allowed a requirement elicitation, which was used for designing PACO-T. In the case study, two Nursing Professors from the Nursing Department of the Federal University of São Carlos planned a learning activity to prepare their students on how to orient caregivers in the community from which the common sense knowledge was collected. In the learning activity, the common sense knowledge was used to call the learners' attention to the way which the population talked about requirements to be a caregiver or about procedures which might be taken while home caring a sick person and to point which learners should emphasize during the orientation were presented (Carvalho et al., 2007; Carvalho et al., 2008a). Table 5 summarizes the support that common sense knowledge can give to teachers during learning activities planning using PACO, identified during the case study.

| Step | Support |
|---|---|
| 1 | To define the learning activity theme. |
|   | To compose the learning activity justification. |
|   | To define the learning activity general objective. |
|   | To define the learning activity specific objectives. |
| 2 | To decide which topics should be approached during the learning activity, so that it can fit to the students' needs. |
|   | To decide the detail degree with which each topic should be approached. |
| 3 | To reach pedagogical issues addressed in Freire's (2003), Freinet's (1993), Ausubel's (1968) and Gagné's (1977) Learning Theories. |
| 4 | To fit the learning tasks to the pedagogical/ methodological references adopted. |
|   | To know how the target group usually study. |
| 5 | To know with which computer tools the target group is familiar. |
| 6 | To compose the learning material. |
| 7 | - |

Table 5. Support offered by common sense knowledge in each PACO's Step

In PACO-T, the target group's common sense knowledge is presented to educators, so that they can assess how people from that target group talk about topics related to the theme of the learning activity being planned and decide which topics they should approach in the learning activity, according to the needs they identify. Through common sense analysis educators can become aware about the learners' level of knowledge, avoiding approaching topics which learners already know and approaching topics which is misunderstood by that learners' profile, since common sense registers myths, believes and misunderstandings of people from whom the knowledge was collected.

The tool uses the semantic networks and the API provided by OMCS-Br to present knowledge related to the context of the learning activity planning, collected from volunteers with the profile of target group. For this purpose, teachers should define the target group's profile at the beginning of the planning so that a suitable ConceptNet can be instantiated and used during it. Figure 9 presents one of PACO-T interfaces with common sense support. The common sense support box can be noticed on the right.

Note that the items are presented in the common sense support box as links. Clicking on the link, the system performs a new search in ConceptNet, using as keyword the text of the link and updates the information in the common sense support box. This allows educators to navigate among the concepts previously presented and to analyze how people with the same profile of their target group talk about related subjects. Concerning the content in Figure 9, by analyzing it the educator can see that one of the motivations of home caring a sick person is to save money. Therefore, s(he) can consider that for her/his target group it is important to know procedures of home caring sick person which are not expensive and approach this theme during the learning activity.

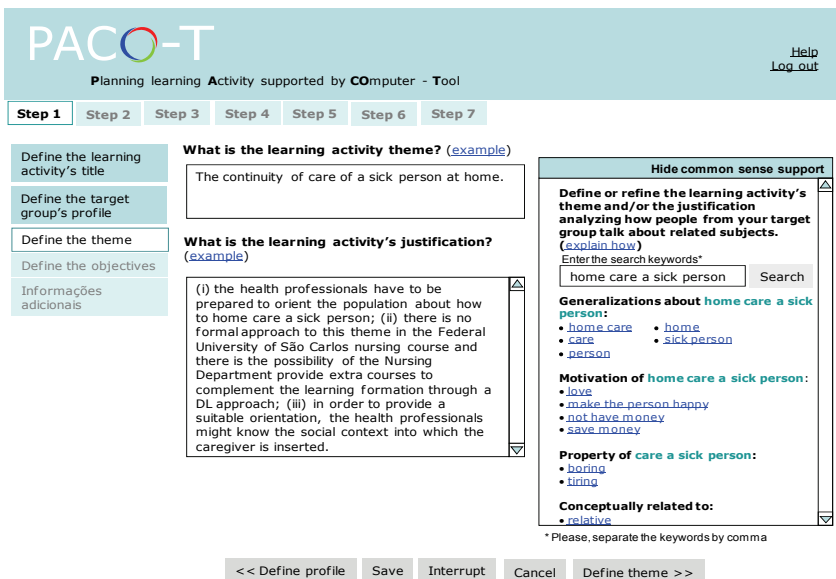

Fig. 9. PACO-T Interface – Step 1: Define the theme

Tools such as PACO-T, which can personalize the information that it is going to present to users, according to some parameters they provide, can be considered an valuable contribution to HCI, especially when the development of adjustable interfaces, which seems to be one of the main issues of HCI in a near future, is taken into account.

### 4.2. Developing systems which infer over cultural issues

Imagine the possibility of developing an agenda system capable of warning someone who tries to set up a meeting with an Indian in a barbecue place that this is not the best place to take an Indian, based on the information "Indians does not eat cow meat, because cows are considered holly in its culture". This is undoubtedly an advance in HCI that will allow the development of systems capable of acting flexibly to different kind of situations based on cultural information.

To support teachers contextualizing considering the educational context, the learning content they are preparing by suggesting topics of interest for the target group through inferences over a common sense knowledge base, an example to illustrate the use of common sense knowledge for the purpose of developing systems which infer over cultural issues. Moreover, when communities of people with common interests are considered like a group of learners of a certain educator in certain school year, it is possible to improve the interaction between educator and learners allowing educators to know about the learners' daily life and to prepare the content they will approach in classroom, considering the learners' cultural context. If a certain concept is explained through metaphors and analogies coming from learners' community the chances of such concept to be understood by the learners are bigger, according to pedagogical principles. Then, developing systems capable of suggesting those examples by inferring over a common sense knowledge base can be helpful. This section presents Cognitor, a computational framework for editing web-like learning content, and a Common Sense-Based On-line Assistant for Training Employees which implement those ideas.

***Cognitor: a framework for editing culture-contextualized learning material***

Cognitor is an authoring tool, based on the e-Learning Pattern Language Cog-Learn (Talarico Neto et al., 2008). Its main objective is to support teachers in designing and editing quality learning material to be delivered electronically to students on the Internet or on CD/DVD. It aims to facilitate the development of accessible and usable learning material, which complies with issues from Pattern and Learning Theories, giving teachers access to common sense knowledge in order to make them able to contextualize their learning materials to their learners' needs (Anacleto et al., 2007a). About using common sense knowledge in the learning process, Cognitor is being prepared to make possible the use of common sense knowledge through all the edition of the learning content, in order to instantiate the Patterns implemented in the tool, to organize the hyper document, to contextualize the content, to adjust the vocabulary used in the content and to fulfill the metadata of such learning objects.

One of the tools available in Cognitor is the implementation of the Cog-Learn Knowledge View Pattern (Talarico Neto et al., 2008), which supports the planning and the automatic generation of learning material navigational structure, by using the technique of Concept Map, proposed by Novak (Novak, 1986). For that purpose, teachers should: (i) enter into the system the Concepts which they want to approach in their learning material and organize them hierarchically; (ii) name the natural relations between concepts, i.e. the relations mapped from the hierarchy previously defined; and (iii) establish and name any other relation between concepts, which they consider important to the content understanding and exploration. See (Anacleto et al., 2007a) for more information on the process of generating a Concept Map in Cognitor, using the Knowledge View Pattern.

When the teacher chooses to use the Knowledge View pattern, Cognitor offers the common sense support to provide her/him with information about the facts that are considered common sense in the domain that s/he is thinking of. So the teacher can complete the Concept list based on that information, decreasing the time on planning the material organization. Figure 10 depicts the first step of Cog-learn Knowledge View Pattern tool, in which teachers are expected to enter into the system the concepts which they want to approach in the content.

In the example, the teacher entered the concept "Health" ("Saúde", in Portuguese), and the system suggested in the box on the right concepts such as "sick person" ("pessoa doente"), "place" ("lugar") and "drug" ("remédio"). Teachers can select one of the suggestion from common sense knowledge and click on the button "<< Include", select a concept in the list of concepts on the right and perform a search in the common sense knowledge base for related concepts through the button "Search >>", or even add another concept that is not in the common sense concepts suggestion list. By using common sense suggestion educators can contextualize the learning content they are preparing to their target group´s needs in the same way it is done in PACO-T.

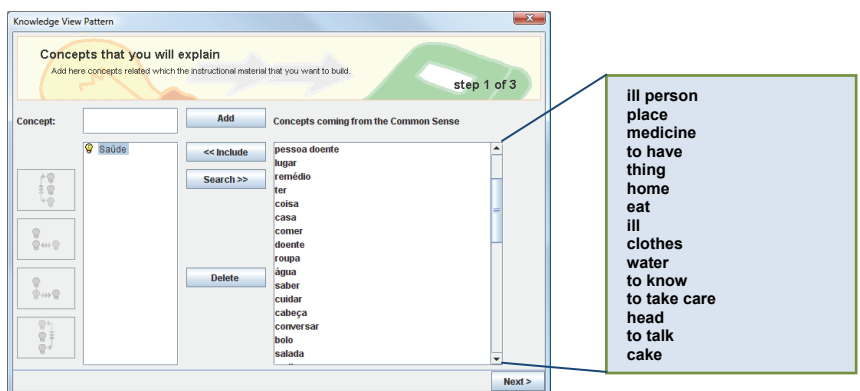

Fig. 10. Knowledge View Pattern First Step

### On-line Help Assistant for Training Employees

The On-line Help Assistant is a web system to support continuing education about workplace safety issues, where the learner can ask a doubt about a certain concept in natural language and the system gives him back the formal definition and some analogies coming from common sense of that employee community to explain the concept (Anacleto et al., 2007b). In order to build the analogies, the system uses the function *get_analogy()* mentioned in section 3.3 of this chapter.

The system is basically a common sense-aided search engine. Its interface is composed for five areas, four of them presented in Figure 11:
1. **Search area**, where users should type the search query;
2. **Definition area,** where definitions for the concepts in the search query are presented;
3. **Analogy area,** where the analogies to explain the concepts are shown;

   4.  **Relations area,** where the relations among the concepts present in the search query and the ones in the ConceptNet are presented; and
   5.  **Related concept area**, where related concepts retrieved from ConceptNet are presented and users can continue their studies by exploring those concepts. For this purpose, learner can use the links offered by the assistant and browse through the concepts.

Users should type the concept they want to get explained and then click on the button "Shown information found". In order to identify the morphological variations in the query, the system uses two techniques. The first one is the expansion technique. It is useful when students use short expressions to perform the search. In this technique, the terms related to the context are retrieved by the function *get_context()* of the ConceptNet API. Then, terms that have the lemmatized expression as a substring are also retrieved. For instance, when a learner provides the expression "fogo" ("fire", in English) in the search area the system will also search for expressions like "fire alarm", "fire place", "fire door" and so on. The second technique is useful especially when large expressions are provided by the students to be searched. In this case phrasal structures, such as noun phrases, verbal phrases and adjective phrases are identified by the system and then the system performs a search for each structure identified. For example, when a student asks the system for results related to the expression "prevenir acidentes no ambiente de trabalho" (in English, "preventing accidents in the workplace"), the expression will be divided in "preventing work accidents", "work accidents", "accidents" and "accident in the workplace". This technique increases the likelihood of getting results from the search, since the terms that are searched are simpler and more likely to be found in the semantic network (Anacleto et al., 2007b).

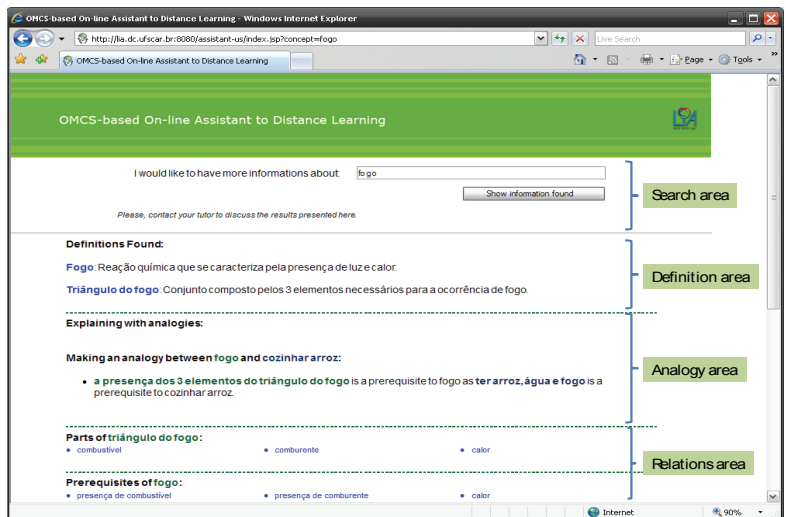

Fig. 11. On-line Help Assistant for Training Employees Interface

### 4.3. Designing systems considering cultural issues

Attributes as attraction, dynamism, level of expertise, confidence, intention, location, social validation and preferences have different weights in different cultures. Consequently, user interface developers face many challenges, such as the design of the content which should be presented for users with different culture profiles. For example, which colors should be used in the design of an interface in order to make users feel comfortable or react in the desired way? Which symbols should be used so that the right message is transmitted?

Despite of the importance of these issues, developers face the almost unreachable tasks of researching culture-dependent requirements and having a contextualized development accepted due to tight budgets, limited agenda, and scarce human resources. In this context, collecting these "world views" and making them available for everyone that wants to develop culture-contextualized user interfaces, can be expensive and laborious.

Taking into account that this kind of information and that the answer for question´s such the ones stated before can be found in people's common sense. Using this kind of knowledge in the system development seem´s to be a alternative for providing cultural contextualization.

Common sense-based interface content design has been tested at LIA in an educational game called "What is it?" (Anacleto et al., 2008a), a card-based quiz game that proposes to work on curricular themes. The game framework is presented in the following.

### *"What is it?": a framework for designing culture-contextualized educational games*

The game framework "What is it?" is a proposal to support teachers in contextualizing the content of quiz-like educational games to the students' local culture, in order to promote a more effective and significant learning. It is divided in two modules: a module for educators, called "Editor's Module" and another for learners, called "Player's Module" (Anacleto et al., 2008a). The cards designed by the educators are presented for the learners in the learner's module, which automatically considers cultural issues when the topics to be worked on are established. So, the content presented to learners is customized according the common sense knowledge shared by people with their profile, once the educator previously established the topics and statements from that learner's community to set up the game (Anacleto et al., 2008a).

**Editor's Module**

The game editor is a seven-step wizard which guides the teacher towards creating game instances, which fit to their pedagogical goals. It implements the principle of developing systems that shows cultural issues for decision-making, presented in section 4.1.

During the card definition, teachers receive the support of common sense knowledge. For that purpose, in the framework editor Step 1 teachers have to define the population profile which should be considered in the search for common sense statements in the knowledge base. In this way, the system guarantees that the statements which are going to be presented to the teacher were gathered from people who have the desired profile to the game instance, i.e. the statements are contextualized to the target group.

In the two next steps the teacher must define two items: (1) the game main theme that nowadays have to be related to one of the six transversal themes defined on the Brazilian curriculum (sexual education, ethics, healthcare, environment, cultural plurality, market and consumers) and (2) the topics, which are specifics subjects related to the theme chosen, to compose the game dice faces (Anacleto et al., 2008a).

The next steps consist of defining the secret words, their synonyms and the set of clues for each secret word. For each secret word defined, it is performed a search on the ConceptNet instantiated at Step1, that increasing the number of words associated with the secret work. The concepts associated with the secret word and their synonyms are presented to teachers as natural language statements and, based on these statements, teachers can compose the card clues. For example, the relation (IsA "aids", "sexually transmitted disease"), found in the ConceptNet-Br, is presented to teachers as "Aids is a(n) sexually transmitted disease".

Then the teacher can (a) select a common sense statements as clues, adding them to the card set of clue; (b) edit a statement to make it suitable to the game purpose; or (c) just ignore the suggestions and compose others clues. It is worth pointing out that the statements edited or composed by the teachers are also stored in the OMCS-Br knowledge base as new common sense statement collected from that teacher.

It is also important to point out the fail-soft approach adopted in the framework. This means that the statements suggested to teachers can be valid or not and the teachers will decide for accepting or not the suggestion. However, the suggestion does not bring any problem to the teachers' task performance. On the contrary, it helps the teachers to finish their task faster and more efficiently.

**Player's Module**

Figure 12 presents an instance of "What is it?" in theme "Sexual Education". That is the interface presented to learners after teachers finish preparing the game cards. To start the game the player should click on the virtual dice, represented in Figure 12 by the letter "S", whose faces represent the topics related to the transversal theme on which the teacher intents to work. In Figure 12, the letter "S" corresponds to the topic "Sexually transmitted diseases". Other topics which can potentially compose the "Sexual Education" theme dice, according to the teachers' needs, are "anatomy and physiology", "behavior" and "contraceptives methods". The letters, which represent the topics, are presented to the player fast and randomly. When the player clicks on the dice it stops and say about which topic the secret word, which should be guessed, is.

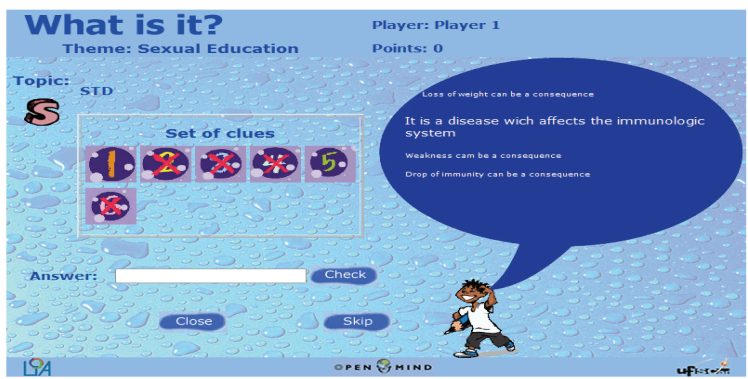

Fig. 12. Player's Module Main Interface

Each topic has a set of cards associated with, which are related to different secret words. These cards are defined by teachers in the game's editor module, using the support of a common sense knowledge base. In addition to that, it is possible to relate a list of synonyms to each secret word. These synonyms are also accepted as expected answers.

The clues play the role of supporting the player to guess which the secret word is. Each card can have a maximum of ten clues which can be selected by the learners by clicking on a number into the "Set of clues" area, which can be seen in Figure 12. After having the topic defined by the dice, a card with clues is presented to the player and, as s(he) selects a clue, it is displayed on the blue balloon. The players can select as many clues as they consider necessary before trying to guess the word.

As the players try to find out the secret word, the system collects common sense knowledge, storing the relation between the word they have suggested and clues that have been already displayed. This collection process is interesting (1) for teachers, who can identify possible misunderstanding by analyzing the answers that learners with the profile of their target group give to a specific set of clues, and, therefore, approach those misunderstandings in classroom to clarify them; and (2) for the OMCS-Br, which will get its knowledge base increased. It is important to point out that the answers provided by the learners, which do not correspond either to the secret word or to a synonym defined by the teacher, are not considered incorrect by the system.

## 5. Conclusion and Future Works

The advent of Web 3.0, claiming for personalization in interactive systems (Lassila & Hendler, 2007), and the need for systems capable of interacting in a more natural way in the future society flooded with computer systems and devices (Harper et al., 2008) show that great advances in HCI should be done.

This chapter presents some contributions of LIA for the future of HCI, defending that using common sense knowledge is a possibility for improving HCI, especially because people assign meaning to their messages based on their common sense and, therefore, the use of this knowledge in developing user interfaces can make them more intuitive to the end-user. Moreover, as common sense knowledge varies from group to group of people, it can be used for developing applications capable of giving different feedback for different target groups, as the applications presented along this chapter illustrate, allowing, in this way, interface personalization taking into account cultural issues.

For the purpose of using common sense knowledge in the development and design of computer systems, it is necessary to provide an architecture that allows it. This chapter presents LIA's approaches for common sense knowledge acquisition, representation and use, as well as for natural language processing, contributing with those ones who intent to get into this challenging world to get started.

Regarding the educational context adopted by LIA's researchers, the approaches presented here go towards one of the grand challenges for engineering in 21st century recent announced by the National Academy of Engineering (NAE) (http://www.nae.edu/nae/naehome.nsf): advance personalized learning (Advance Personalized Learning, 2008). Using intelligent internet systems, capable of inferring over the learning contexts which educators want to approach and presenting related themes, topics and pieces of information retrieved from is one of the possible approaches mentioned by NAE.

As future work it is proposed to perform user tests on the applications presented along this chapter, in order to check their usability and, consequently, the users' satisfaction in using such kind of application.

## 6. Acknowledgements

We thank FAPESP and CAPES for financial support. We also thank all volunteers who have been collaborating with the Project OMCS-Br and all LIA's researchers, who have been giving valuable contributions towards the project.